\documentstyle[12pt, aasms4, epsfig]{article}

\def\etal{{\it et al.}}
\def\ie{{\it i.e.}}
\def\eg{{\it e.g.}}

\def\gap{\hbox{${_{\displaystyle>}\atop^{\displaystyle\sim}}$}}

\def\del{{\mathbf \nabla}}

\begin{document}

\title{Quaking Neutron Stars}

\author{Lucia M. Franco\altaffilmark{1}}
\affil{University of Chicago, 5640 S. Ellis Ave., Chicago IL 60637;
lucia@oddjob.uchicago.edu}
\author{Bennett Link\altaffilmark{1}}
\affil{Montana State University, Department of Physics, Bozeman MT
59717; blink@dante.physics.montana.edu}
\altaffiltext{1}{Also Los Alamos National Laboratory}
\and
\author{Richard I. Epstein}
\affil{Los Alamos National Laboratory, Mail Stop D436, Los Alamos, NM
87545; epstein@lanl.gov}

\begin{abstract}

Gravitational, magnetic and superfluid forces can stress the crust of
an evolving neutron star.  Fracture of the crust under these stresses
could affect the star's spin evolution and generate high-energy
emission. We study the growth of strain in the crust of a spinning
down, magnetized neutron star and examine the initiation of crust
cracking (a {\em starquake}). In preliminary work (Link, Franco \&
Epstein 1998), we studied a homogeneous model of a neutron star.  Here
we extend this work by considering a more realistic model of a solid,
homogeneous crust afloat on a liquid core. In the limits of
astrophysical interest, our new results qualitatively agree with those
from the simpler model: the stellar crust fractures under shear stress
at the rotational equator, matter moves to higher latitudes and the
star's oblateness is reduced.  Magnetic stresses favor faults directed
toward the magnetic poles. Thus our previous conclusions concerning
the star's spin response still hold; namely, asymmetric
redistribution of matter excites damped precession which could
ultimately lead to an increase in the spin-down torque.
Starquakes associated with glitches could explain the permanent {\em
offsets} in period derivative observed to follow glitches in at least
three pulsars.

\end{abstract}


\section{Introduction}

Stresses developing in the crust of an evolving neutron star could 
fracture the crust (\ie\ a starquake), possibly affecting the star's
spin evolution and generating high-energy emission. Spin down (\eg, in
isolated pulsars; \cite{starquakes}) or spin up (\eg, in accreting neutron
stars) changes the equilibrium shape of the star, building stresses. In
``magnetars'', decay of the superstrong field ($B\gap 10^{14}$ G) could
break the crust and produce episodes of intense gamma-ray emission
(\cite{td}; \cite{tb}). Differential rotation between the crust and the
interior neutron superfluid may also stress the crust to its breaking point
(\cite{ruderman76}). 

Starquake-induced rearrangement of the star's mass distribution
changes the star's moment of inertia causing precession and polar
wandering (Link, Franco \& Epstein 1998; hereafter LFE).  Anomalous
spin behavior might therefore signal the occurrence of a
starquake. Such evidence for crust cracking may already exist in
several isolated pulsars. Permanent {\em offsets} in period derivative
following glitches have been observed in the Crab pulsar (\cite{LPS}),
PSR1830-08 and probably PSR0355+54 (\cite{sl96}). The observed offsets
all have the same sign, and correspond to increases in spin-down
rate. In LFE we interpreted these offsets as due to permanent
increases in the torque acting on the neutron star. We showed that
cracking and readjustment of the stellar crust in response to the
star's slow down can increase the angle between the star's
spin and magnetic axes, leading to an increase in the spin-down torque
in some models of pulsar spin down.

Other evidence for cracking of the neutron star crust can be found in
the Soft Gamma Repeaters, thought to be strongly magnetized neutron
stars ($B\gap 10^{14}$ G), or {\em magnetars}.  Unlike radio pulsars
which have weaker magnetic fields and are powered by rotation,
magnetars are thought to be powered by their intense magnetic
fields. Duncan \& Thompson (1994; also Thompson \& Duncan 1995; 1996)
have suggested that SGR outbursts represent cracking of the crust by
magnetic stresses. Evidence in favor of this hypothesis is the
striking statistical similarities between bursts in SGRs and
earthquakes (\cite{cheng}; \cite{1900+14}). For both
phenomena the energy released per event obeys similar power-law
scalings and the waiting times between events are strongly correlated
with one another.

The potential importance of violent crust dynamics in neutron star
evolution motivates this detailed study of spin-down induced
starquakes. The equilibrium shape for a spinning star is an oblate
spheroid which becomes more spherical as the star slows down.  The
liquid core is able to change its shape smoothly as the star slows
down. In contrast, the solid crust is strained as it evolves through a
continuous series of equilbrium configurations, and eventually will
crack if it is brittle. In \cite{lfe98}, we modeled the neutron star
as a homogeneous, self-gravitating, elastic sphere and followed the
evolution of strain in its crust as the star spins down. From our
calculations we developed the following picture of starquakes. Crust
cracking occurs as equatorial material shears under the compressive
forces arising from the star's decreasing circumference.  The star's
oblateness suddenly decreases as matter moves to higher latitudes
along a slip or {\em fault} plane that is perpendicular to the stellar
surface and crosses the equator at an angle of $\sim 30-45^\circ$ (see
Fig. 6). Since magnetic stresses suppress shearing near the magnetic
poles and across the field lines, starquakes likely originate near the
two points on the equator farthest from the magnetic poles and
propagate preferentially toward the magnetic poles. The matter
redistribution breaks the rotational symmetry of the star, causing the
star to precess.  Damping of the precession eventually restores
alignment between the angular velocity and angular momentum,
ultimately {\em increasing} the angle between the spin axis and the
magnetic moment. In some models of pulsar spin-down (\eg , the
vacuum-dipole model) an increase in the alignment angle between the
rotation and magnetic axes increases the spin-down torque.  The
magnitude of the starquake-induced changes in the alignment angle are
consistent with the changes in the spin-down rates seen in the Crab
pulsar (\cite{lfe98}). The observations of spin-down offsets in
coincidence with glitches indicates some connection between structural
readjustments and glitches. For a recent discussion of this
possibility, see Link \& Epstein (1996).

In this paper we extend the analysis to the more realistic case of a
brittle shell floating on a liquid core. We find that the stresses
induced by the star's spin down produce similar behavior to that found
in LFE. Crust cracking associated with other sources of stress, such as
magnetic or superfluid stresses, will be considered in future work.
The organization of this paper is as follows. Section
\ref{phys-pict} presents a qualitative description of the growth
and release of strain in the crust of an idealized, spinning-down
neutron star. In \S \ref{the-math}, we calculate the response of the
crust's shape as it spins down, treating the star as a homogeneous
solid afloat on a liquid core. In \S \ref{thickness}, we explore
the cracking and matter flow for a model with realistic crust
thickness.  In \S \ref{b-field} we discuss the effects of a magnetic
field anchored in the crust. Section
\ref{summary} summarizes our results.

\section{Growth and Release of Strain in a Spinning-down Star}
\label{phys-pict}

Our analysis is based on the assumption that the neutron star crust is
brittle, and therefore cracks when sufficiently
strained\footnote{Experience with terrestrial materials shows that
material properties at high pressure can differ from those at low
pressure. For example, terrestrial materials typically become ductile
when subjected to pressures in excess of their shear moduli.
Nevertheless, deep-focus earthquakes are known to originate from regions
of very high pressure (\cite{gh95}). These faults are thought to be
triggered by densification phase changes; small regions of higher
density nucleate as the material is stressed, and act as a lubricant for
shearing motion. Analogous processes might occur in the high-pressure
material of the neutron star. E. Ramirez-Ruiz and R. I. Epstein are
examining whether the phase transition from spherical nuclei to rod-like
nuclei may facilitate starquakes.}. We consider a neutron star spinning
with angular velocity $\Omega$ which undergoes a change $-\delta\Omega$
in its rotation rate where $0 < \delta\Omega \ll \Omega$. Our goal is to
describe the effects on the star's crust resulting from this change in
rotation rate. In equilibrium, the star has a spheroidal shape with an
equatorial bulge of relative size $\sim \left(\Omega R / v_k\right)^2$
where $R$ is the radius of the non-rotating configuration and $v_k$ is
the Keplerian velocity. As the star spins down, the fluid interior
readjusts its shape continuously, following (in the homogeneous case) a
sequence of Maclaurin spheroids (see, \eg, Chandrasekhar, S. 1987;
p.77). As the solid crust attempts to readjust its shape, however,
strain builds. Material originally at ${\mathbf r}$ is displaced to
${\mathbf r} + {\mathbf u}({\mathbf r})$, where ${\mathbf u}({\mathbf
r})$ is the {\em displacement field}.  The local distortion of the solid
is given by the {\em strain tensor} (see, \eg, Landau \& Lifshitz
1959; p.3)
\begin{equation}
\label{strain_tensor}
u_{ij} = {1\over 2}\left ({\partial u_i\over\partial x_j} +
                          {\partial u_j\over\partial x_i}\right ).
\end{equation}
In a local coordinate system in which this matrix is diagonal, the
eigenvalues $\epsilon_{ii}$ represent compression ($\epsilon_{ii}<0$)
or dilation ($\epsilon_{ii}>0$) along the respective axes.  Let the
largest and smallest eigenvalues be $\epsilon_l$ and $\epsilon_s$,
respectively. The {\em strain angle} is $\epsilon_l - \epsilon_s$, and
we refer to the plane containing the corresponding principal axes as
the {\em stress plane}. In Fig. 1 we show schematically a block of
matter that is compressed along the $y$ direction and allowed to
expand along the $z$ direction. The stress plane then is parallel to
the $y-z$ plane.

The elastic limit of the matter is reached first in the regions of the
crust where the strain angle is maximum. There the material cracks,
forming a fault plane perpendicular to the stress plane
(Fig. 1). Matter slips along the fault plane and redistributes in a
way that reduces the equatorial circumference, thus releasing the
strain. The fracture geometry depends on where the strain angle is
maximum and how its value changes over the crust. We now turn to a
calculation of the spin-down induced strain in the neutron star crust.

\section{Matter Redistribution in a Spinning-down Star}
\label{the-math}

To explore the accumulation and release of strain in a slowing
neutron star, we model the star as a two-component homogeneous
spheroid; a self-gravitating core of incompressible liquid on which
rests a crust of uniform density. The non-rotating configuration
is a sphere of core radius $R'$ and total radius $R$.

\subsection{Equilibrium Equations}
\label{set-up}

A spinning, self-gravitating object has an equilibrium
configuration given by 
\begin{equation}
\del\cdot{\mathbf \mathsf{T}} - \rho\del\phi = {\mathbf F},
\label{equil}
\end{equation}
where ${\mathbf \mathsf{T}}$ is the material stress tensor, $\phi$ is
the gravitational potential per unit mass and ${\mathbf F}$ is the
centrifugal force density associated with the rotation. For rigid
material, strain develops as the force is changed. The strain in the
material is given by the strain tensor of
eq. (\ref{strain_tensor}).  For incompressible matter we have
\begin{equation}
\del\cdot{\mathbf u}= u_{ii}=0,
\label{incompress}
\end{equation}
where repeated indices indicate sums.

For sufficiently small strains, the material can be described using
Hooke's Law in terms of the stress tensor (see, \eg, Landau \&
Lifshitz 1959, p. 11).
\begin{eqnarray}
{\mathbf \mathsf{T}}_{ij} & = & -p \delta_{ij} + \sigma_{ij} \label{Tij}\\
\sigma_{ij} & = & 2\mu u_{ij},
\label{sigmaij}
\end{eqnarray}
where $p$ is the isotropic pressure, $\mu$ is the elastic shear modulus,
$\sigma_{ij}$ represents the contribution of the material's rigidity to
the stress, and we have applied eq. (\ref{incompress}).

We consider an initial configuration in which the material is under
great internal pressure but unstrained ($u_{ij}=0$). The equation
of hydrostatic equilibrium is then
\begin{equation}
-\del p_0 - \rho\del\phi_0 = {\mathbf F}_0,
\label{unperturbed}
\end{equation}
and, in the limit of slow rotation, $R\Omega/v_k \ll 1$, $F_0$ is small
compared to both $\nabla p_0$ and $\rho \nabla \phi_0$.

To derive the condition for hydrostatic equilibrium in the new
configuration produced by the change in ${\mathbf F}$, we consider a
Lagrangian perturbation of the initial state.  A Lagrangian perturbation
$\Delta$ is evaluated moving with the matter, and is related to an
Eulerian perturbation $\delta$ through
\begin{equation}
\Delta = \delta + {\mathbf u}\cdot\del.
\end{equation}
A Lagrangian perturbation commutes with spatial derivatives as (see,
\eg, Shapiro \& Teukolsky 1983; p. 131)
\begin{equation}
\Delta{\partial\over\partial x_i} = {\partial\over\partial x_i} \Delta
                     - {\partial u_j\over\partial x_i}\nabla_j.
\end{equation}

We obtain the new equilibrium by taking a Lagrangian perturbation of
eq. (\ref{equil}), \ie,
\begin{equation}
\Delta\left [\del\cdot{\mathbf \mathsf{T}} - \rho\del\phi = {\mathbf
F}\right ].
\end{equation}
Using eqs. (\ref{strain_tensor}) through
(\ref{unperturbed}), we obtain, 
\begin{equation}
- \del\Delta p - \rho c_t^2\,\del\times\del\times{\mathbf u} -
\rho\del\delta\phi =
     \delta{\mathbf F} - \del ({\mathbf u}\cdot\del p_0)
     + ({\mathbf u}\cdot \del) {\mathbf F_0}
     - {\mathbf u}_k {\partial F_k\over\partial x_i},
\label{perturbed}
\end{equation}
where we have neglected terms of order ${\mathbf u}^2$ and
$c_t=\sqrt{\mu/\rho}$ is the transverse sound speed, which we take to be
constant throughout the solid.  For the case of
slow rotation, $R\Omega/v_k \ll 1$, the last two terms in eq.
(\ref{perturbed}) are smaller than the leading terms by a factor
$\sim(\Omega R /v_k)^2$, and we neglect them.

We express the Eulerian change in centrifugal force by means of a
scalar potential $\chi$ defined as
\begin{equation}
\delta {\mathbf F}=\rho \del\chi.
\end{equation}
Introducing a second potential $h$,
\begin{equation}
c_t^2\,h \equiv -{1\over\rho}\Delta p - \delta\phi - \chi
+ {1\over\rho}{\mathbf u}\cdot\del p_0,
\label{hdef}
\end{equation}
we rewrite eq. (\ref{perturbed}) as
\begin{equation}
\del\times\del\times{\mathbf u} = \del h.
\label{potentialeqn}
\label{displacement}
\end{equation}
Taking the divergence of this equation, we see that the potential satisfies
\begin{equation}
\nabla^2 h = 0.
\label{heqn}
\end{equation}
The displacement ${\mathbf u}$ is obtained by solving eqs.
(\ref{potentialeqn}) and (\ref{heqn}) with (\ref{hdef}) subject to boundary
conditions.

\subsection{Displacement Field of the Crust}
\label{solving-u}

In a spherical coordinate system $(r,\theta)$ with ${\mathbf \Omega}$
directed along the $z$-axis, the centrifugal potential associated with
a change in rotation rate $\delta\Omega$ is
\begin{equation}
\chi \equiv {2\over 3}\Omega\delta\Omega r^2 ( 1 - P_2 (\theta)),
\label{chi}
\end{equation}
where $P_2= (3 \cos^2\theta -1)/2 $ is the second Legendre polynomial.
The $P_2 (\theta)$ symmetry of $\chi$ suggests a solution
to eq. (\ref{heqn}) of the form
\begin{equation}
h (r, \theta) = \left (A r^2 + {B\over r^3}\right ) P_2 (\theta) + C,
\label{hsoln}
\end{equation}
where $A$, $B$ and $C$ are constants. For the displacement field, we
seek a solution of the form
\begin{eqnarray}
u_r (r, \theta) = f (r) P_2 (\theta) \label{uassumedr}\\
u_\theta (r, \theta) = g (r) {dP_2 (\theta)\over d\theta}
\label{uassumedt}
\end{eqnarray}
where $f(r)$ and $g(r)$ are functions of depth in the star only.
The solution to eqs. (\ref{potentialeqn}) and (\ref{heqn}), with
(\ref{hsoln}), and subject to condition (\ref{incompress}), is
\begin{eqnarray}
u_r (r, \theta) = \left (a r - {A\over 7} r^3 - {B\over 2 r^2} +
        {b\over r^4}\right ) P_2 (\theta) \label{ur} \\
u_\theta (r, \theta) = \left ( {a r\over 2} - {5\over 42} A r^3 -
        {b\over 3 r^4} \right ) {dP_2 (\theta)\over d\theta},
\label{ut}
\end{eqnarray}
where $a$ and $b$ are additional constants. We require four boundary
conditions to specify
the four coefficients.

\subsection{Boundary Conditions}
\label{bdry}

At the outer boundary the crust encounters a vacuum, while at the inner
boundary it is in contact with the liquid core. Since neither of these
media  support traction, two boundary conditions follow from
requiring that no shear exist at each boundary, \ie,
\begin{equation}
\sigma_{r\theta} = 0 \qquad {\mathnormal at} \qquad r=R,R^\prime,
\end{equation}
or,
\begin{equation}
{\partial u_\theta\over\partial r} - {u_\theta\over r} +
           {1\over r}{\partial u_r\over\partial\theta} = 0
                    \qquad {\mathnormal at} \qquad r=R,R^\prime.
\end{equation}
In terms of the coefficients
\begin{equation}
a - {8\over 21}A R^2 - {B\over 2R^3} + {8\over 3}{b\over R^5} = 0
\label{bc1}
\end{equation}
\begin{equation}
a  - {8\over 21}A R^{\prime 2} - {B\over 2R^{\prime 3}} + {8\over
3}{b\over R^{\prime 5}} = 0.
\label{bc2}
\end{equation}
Two other boundary conditions follow from requiring that the material
be in mechanical equilibrium at the boundaries. This requirement is
satisfied by ensuring that
\begin{equation}
{\mathbf {\mathsf T}}_{rr} = - p + \sigma_{rr},
\label{Tjump}
\end{equation}
is continuous at $r=R,R^\prime$. Hence,
\begin{equation}
\left [-\Delta p  + 2\rho c_t^2 {\partial u_r\over\partial r}\right
]_{r=R-\varepsilon} = 0,
\label{pjumpsurf}
\end{equation}
\begin{equation}
\ \qquad\qquad\left [-\Delta p  +
2\rho c_t^2 {\partial u_r\over\partial r}\right ]_{r=R^\prime+\varepsilon}
=
\left [-\Delta p\right ]_{r=R^\prime-\varepsilon},
\label{pjumpcore}
\end{equation}
where $\varepsilon$ is an infinitesimal positive distance. We have
required $\Delta p=0$ just above the star and have taken $c_t=0$ in the
liquid core.  These jump conditions specify the material stress required to
maintain a pressure difference across a boundary.

The local pressure change follows from eq. (\ref{hdef})
\begin{equation}
\Delta p = -\rho c_t^2 h - \rho\,\delta\phi - \rho\chi +
               {\mathbf u}\cdot\del p_0.
\label{pchange}
\end{equation}
The initial, unstrained configuration is that of
a Maclaurin spheroid whose shape, for small eccentricity $e$, can be
written
\begin{equation}
R (\theta) = R \left[1 - \frac{1}{3} e^2 P_2(\theta) \right] \equiv
R + \epsilon(\theta).
\end{equation}
The gravitational potential for a slowly-rotating Maclaurin spheroid
is (Shapiro \& Teukolsky 1983, {\sl Black Holes, White Dwarfs and
Neutron Stars}, p. 169)
\begin{equation}
\phi = -\pi G\rho\left [2R^2 - {2\over 3}r^2
           + {4\over 5} {r^2\over R}\epsilon (\theta)\right ].
\label{phi}
\end{equation}
Since the crust of a neutron star contains only $\sim 1$\% of the star's
mass, we neglect the crust's effect on the gravitational potential. As
the star spins down,
$\epsilon$ changes by
$u_r(R,\theta)$. Hence, the {\em Eulerian} change in the potential is
\begin{equation}
\delta\phi = -{4\over 5}\pi G\rho {r^2\over R} u_r(R,\theta).
\label{deltaphi}
\end{equation}
For slow rotation, $\del p_0$ is nearly radial, and we may approximate the
last term in eq. (\ref{pchange}) as
\begin{equation}
{\mathbf u}\cdot\del p_0 \simeq - \rho u_r(r,\theta)\del\phi_0 \simeq
                    -{4\over 3}\pi G\rho^2 r u_r (r, \theta),
\label{term}
\end{equation}
where we used the spherical (non-rotating) gravitational potential in
the last step, which is justified in the limit of slow rotation.

Combining eqs. (\ref{pchange}), (\ref{deltaphi}), (\ref{term}), and
(\ref{chi}) through (\ref{uassumedr}), the boundary condition
(\ref{pjumpsurf}) at the surface yields
\begin{equation}
-2 f^\prime (R) - {2\over 5}{v_k^2\over c_t^2}{f (R)\over R} +
{2\over 3}{\Omega\delta\Omega R^2\over c_t^2} = A R^2 + {B\over R^3},
\label{bc3}
\end{equation}
where $f(R)$ is defined by eq. (\ref{ur}) and
$v_k^2\equiv 4\pi G R^2 \rho/3$.

To match the solutions at the crust-core boundary, we require an
expression for the pressure change in the core. We obtain it from eq.
(\ref{perturbed})
setting $c_t = 0$ since a liquid cannot support shear. We have
\begin{equation}
\del \left [-{\Delta p\over\rho} - \delta\phi
+ {1\over\rho}{\mathbf u}\cdot\del p_0 - \chi\right ] = 0.
\end{equation}
Hence,
\begin{equation}
-{\Delta p\over\rho} - \delta\phi
+ {1\over\rho}{\mathbf u}\cdot\del p_0 - \chi = {\rm constant}
\label{hcore}
\end{equation}
in the core.
The pressure change in the shell obeys
\begin{equation}
-{\Delta p\over\rho} - \delta\phi
+ {1\over\rho}{\mathbf u}\cdot\del p_0 - \chi =
c_t^2 \left [A r^2 + {B\over r^3}\right ] P_2 (\theta) + C.
\label{hshell}
\end{equation}
Thus, the boundary condition (\ref{pjumpcore}) at the crust-core boundary,
using
eqs. (\ref{hcore}) and (\ref{hshell}), gives
\begin{equation}
\left [{\partial u_r\over\partial r}\right ]_{r=R^\prime+\varepsilon} =
-{1\over 2}\left [
AR^{\prime 2} + {B\over R^{\prime 3}}\right ] P_2 (\theta) + {\rm
constant}.
\end{equation}
>From the form assumed for $u_r$, we see that the constant is zero. We
finally obtain
\begin{equation}
f^\prime (R^\prime) =  -{1\over 2}\left [AR^{\prime 2} + {B\over
R^{\prime 3}}\right ].
\label{bc4}
\end{equation}
Eqs. (\ref{ur}), (\ref{ut}), and the boundary conditions (\ref{bc1}),
(\ref{bc2}), (\ref{bc3}) and (\ref{bc4}) complete the description of the
displacement field driven by a change in rotation rate $-\delta\Omega$.

Recent calculations with realistic equations of state suggest that
neutron stars have crusts of thickness $0.05$ -- $0.08R$
(\cite{ns-eos}; \cite{ns-eos-crust}); in the following, we take $t
\equiv R-R'=0.05R$. The corresponding displacement field is shown in
Fig. 2. From the displacement field we can calculate the strain
distribution resulting from spin down. The strain tensor
(\ref{strain_tensor}) can be diagonalized to obtain the strain angle
and the planes along which the material can shear. The geometry
of crust cracking can be obtained from an analysis of the
distribution of strain angles over the crust, to which we now turn.

\section{Cracking the Crust}
\label{thickness}

As the star spins down and the strain increases, the crust breaks when
the local strain angle reaches a critical value. Fig. 3 shows
contours of constant strain angle through the crust; the strain
angle is maximum at the crust-core boundary, in the equatorial
plane. Fig. 4 shows the three eigenvalues of the strain tensor at the
inner boundary of the crust plotted as a function of latitude. The
strain angle, shown by the darker curve, shows discontinuous
derivatives where the maximum and minimum eigenvalues change. The
dashed curve corresponds to the strain angle for a homogeneous elastic
sphere model (\cite{lfe98}) at the same depth for comparison. In
the equatorial plane ($\theta = \pi /2$), the stress plane is the
$\theta\phi$--plane. An element of matter on the equatorial plane
is compressed by adjacent elements on the equator and expands in the
$r$ and $\theta$ directions. Material compressed in this way shears
along one of two planes that takes an angle of $\gamma\sim
30-45^\circ$ with respect to the compressive force (\cite{gh95}; see
Fig. 1). Once the crust breaks, matter moves along faults $F$ or
$F'$, inclined at an angle $\gamma$ to the rotational equator (see
Fig. 6). Since the strain field has azimuthal symmetry, the break is
equally likely to begin at all points on a ring corresponding to the
intersection of the crust core boundary with the equatorial plane. The
inclusion of magnetic stresses, discussed below, breaks this symmetry.

The propagation of the crack is a nonlinear process; cracking affects
the strain field, which in turn affects the propagation of the
crack. In general, cracking increases the stress on unbroken
neighboring material (\cite{hertzberg96}), causing the fault plane to
grow in both width and length. Hence, the crack begins at the base of
the crust in the equatorial plane, propagates toward the surface, and
continues to grow up to some latitude.  When the cracking ends, the
strain field is everywhere subcritical. The width and length of the
fault plane cannot be determined from the linear analysis given here.

In this treatment, we have ignored the fact that the neutron star crust
is stratified. Inspection of eq. [\ref{perturbed}] shows that the most
important material property controlling the development of strain is
$c_t$; in particular, where $c_t$ is small, the strain must be
large. Strohmayer (\etal\ 1991) calculated $c_t$ as a function of
density in the neutron star crust and found that it varies by a factor
of $\sim 10^2$. Cracking of the crust is favored
in regions where $c_t$ is small, possibly causing  the quake epicenter to
be somewhat higher than the base of the crust. We expect that the
effects of density stratification will not affect the overall geometry of
crust cracking described here.

\section{Magnetic stresses}
\label{b-field}

We now examine the effects of the magnetic field that is anchored to
the crust of the neutron star. The displacements induced by the spin
down distort the magnetic field, generating magnetic stresses. If the
magnetic field is not precisely aligned with the rotation axis, the
magnetic stresses will break the azimuthal symmetry of the strain
distribution in the crust.

A complete description of the strain field that develops in magnetized
material is obtained by solving the elasticity equations and Maxwell's
equations self-consistently, subject to boundary conditions at the
star's surface. The surface boundary conditions, however, are complex to
handle; in a realistic neutron star they depend on the exact form of the
density gradients near the surface. To qualitatively illustrate the
influence of the star's magnetic field on where starquakes originate and
how they propagate, we consider the simpler problem of the magnetic
fields effect on the material strain in an unbounded medium.

In the absence of the field, the equilibrium state is given by
\begin{equation}
{\partial\sigma_{ij}\over\partial x_i} + F_j = 0,
\end{equation}
where $F_j$ represents gravitational and centrifugal forces. The presence
of a magnetic field introduces several forces. First, the unperturbed field
can exert a force on the crust. For this study, we assume that the initial
field
$\mathbf{B}$ is dipolar and hence force free; that is, the unperturbed Maxwell
stress tensor ${\mathbf \mathsf{T}}_{ij}$ obeys  $\partial{\mathbf \mathsf{T}}_{ij}/\partial
x_i=0$ in the crust.  Next, the displacement field in the crust  distorts
the field by $\delta\mathbf{B}$, generating a perturbation
$\delta{\mathbf \mathsf{T}}_{ij}$ to
the Maxwell stress tensor.\footnote{The
ratio of magnetic to material stresses is $\sim \beta \equiv
(v_A/c_t)^2$ where $v_A$ is the Alfv\`{e}n speed and $c_t$ is the
transverse sound speed. In the outer crust, we have $\beta = 8 \times
10^{-4}
\left(B/10^{12}\ \rm{G} \right)^2 \left(\rho/10^{10}\ {\rm g\ cm}^{-3}
\right)^{-1} \left(c_t/10^8\ {\rm m\ s}^{-1}\right)^{-2}$ and we are
justified in treating the magnetic effects perturbatively.} The new equilibrium is given by
\begin{equation}
{\partial\over\partial x_i}(\delta\sigma_{ij} + \delta{\mathbf \mathsf{T}}_{ij}) = 0
\end{equation}
where  $\delta\sigma_{ij}$ is the magnetically induce change in the material
stress tensor. 
For an infinite medium (or a finite one in which boundary effects are
unimportant), the solution is
$\delta\sigma_{ij}=-\delta{\mathbf \mathsf{T}}_{ij}$.

Magnetic stresses modify the displacement $\mathbf{u}$ by
$\delta\mathbf{u}$. For a magnetic field that is sufficiently small that
this correction to the displacement field is small ($\vert \delta {\mathbf
u}\vert \ll
\vert
{\mathbf u} \vert$), the perturbation to the Maxwell stress tensor is
\begin{equation}
\delta{\mathbf \mathsf{T}}_{ij} = {1\over 4\pi}\left [\delta B_i B_j +
        B_i\delta B_j - B_{k}\delta B_{k}\delta_{ij}\right ],
\end{equation}
with
\begin{equation}
\label{delta-B}
\delta\mathbf{B} = \nabla\times (\mathbf{u}\times \mathbf{B}).
\end{equation}
Here $\mathbf{u}$ is the displacement field in the absence of the magnetic
field.
For incompressible matter, the correction $\delta u_{ij}$ to the
strain tensor is
\begin{equation}
\delta u_{ij} = {1\over 2\mu}\left (\delta\sigma_{ij} - {1\over 3}
              \delta\sigma_{ll}\right ),
\end{equation}
where $\mu$ is the shear modulus. The eigenvalues of this tensor give
the correction to the strain angle due to the induced magnetic forces.
Using the displacement field of eqs. (\ref{ur}) and (\ref{ut}) in
eq. (\ref{delta-B}), we obtain an estimate of the corrections to the
strain angle on the star's rotational equator; these are shown in
Fig. 5. 

The magnetic field affects our picture of crust cracking in two
ways. First, the existence of the magnetic field strengthens the
material, making it harder to shear. The material is weaker and hence
most likely to shear where the magnetic field is weakest. Consider a
star with a dipolar magnetic field inclined at an angle $\alpha$ to the
rotation axis. In this case the material may begin shearing at the two
points on the rotational equator that are farthest away from the
magnetic poles as shown in the schematic of Fig. 6. Second, as the
material attempts to move along the fault planes, the magnetic field
opposes displacements that cross field lines and hence favors matter
motion long fault $F$ over fault $F'$. This motion moves matter closer
to the magnetic poles. This picture is essentially the same as found for
a simpler homogeneous sphere model (\cite{lfe98}).

\section{Discussion}
\label{summary}

We have examined the evolution of strain in the crust of an idealized
spinning-down neutron star and the initiation of starquakes as the
material reaches critical strain.  We modeled the neutron star as a
self-gravitating core of incompressible liquid that supports a brittle
crust. Crust cracking occurs as material shears under the compressive
forces arising from the star's decreasing circumference.  The star
initially cracks near the star's rotational equator, at the base of the
crust. The shearing motion along the fault decreases the oblateness of the
star and pushes matter to higher latitudes.
Magnetic stresses suppress shearing near the magnetic poles and 
 across the field lines.  Starquakes thus 
originate near the two points on the equator farthest from the magnetic
poles and propagate toward the magnetic poles. Our conclusion is
qualitatively the same as found in our simpler homogeneous spherical model
(\cite{lfe98}).

As material slides along a fault directed toward the magnetic poles,
the star's principal axis of inertia shifts {\em away} from the
magnetic axis, exciting precession. Eventually, damping restores
alignment between the principal axis of inertia and the angular
momentum axis, increasing the angle between the rotation and magnetic
axes. In some models of pulsar spin-down, \eg, the magnetic dipole
model, this growth in the alignment angle increases the spin-down
torque. We might, therefore, expect to see a tendency of alignment
angle to increase with age in young pulsars. In older pulsars,
however, Tauris \& Manchester (1998), find no relationship between
magnetic alignment angle and spin-down torque.

One possible observational signature from a starquake is a change of the
pulse profile. A jump in the alignment angle $\alpha$ changes the
duration of the line of sight's traverse through the pulse emission
cone. In the Crab pulsar, the relative change in pulsed flux could be of
order 1\% (Link \& Epstein 1997). Additionally, the energy released in a
starquake could produce an observable enhancement of the star's
luminosity. The energy release is of order $E\sim 10^{42} (\mu/10^{31}\
\mbox{erg cm}^{-3})(d/10^5\ {\rm cm})^3(\theta_c/10^{-2})^2$ ergs, where
$\mu$ is the average shear modulus along a fault of length $d$ and
$\theta_c$ is the strain angle at which fracture occurs. Some fraction
of the seismic energy will be damped, heating the crust. Eventually, a
soft x-ray thermal wave will emerge from the star's surface. For
localized deposition of $10^{42}$ ergs at a density of $10^{14}$ g
cm$^{-3}$, Tang (1999) calculates a 5\% enhancement of the star's
luminosity occurring about 3 yr after the quake for a star with an
internal temperature of $10^7$ K and a stiff equation of state. For
deposition at $10^{13}$ g cm$^{-3}$ in a star with an internal
temperature of $10^8$ K, the enhancement is $\sim 22$\%. The Chandra
X-ray Observatory should be capable of detecting this enhancement in a
closer source, such as the Vela pulsar.

A starquake could drive non-thermal emission as well.  Seismic energy
that reaches the stellar surface couples to Alfv\'en waves which
propagate into the magnetosphere. Conversion of Alfv\'en waves into
$\gamma$-rays could occur if the waves are charge-starved or if their
amplitudes are comparable to the background field strength of the
magnetosphere (Blaes \etal\ 1989).

\acknowledgements

We thank A. Olinto for valuable discussions and C. Miller for a
critical reading of the manuscript. This work was performed under the
auspices of the U.S.  Department of Energy, and was supported in part
by NASA EPSCoR Grant \#291748, NASA ATP grant \# NAG 53688, by IGPP
at LANL and by the Center for Thermonuclear Flashes at the University
of Chicago.

\newpage

\begin{figure}
\plotone{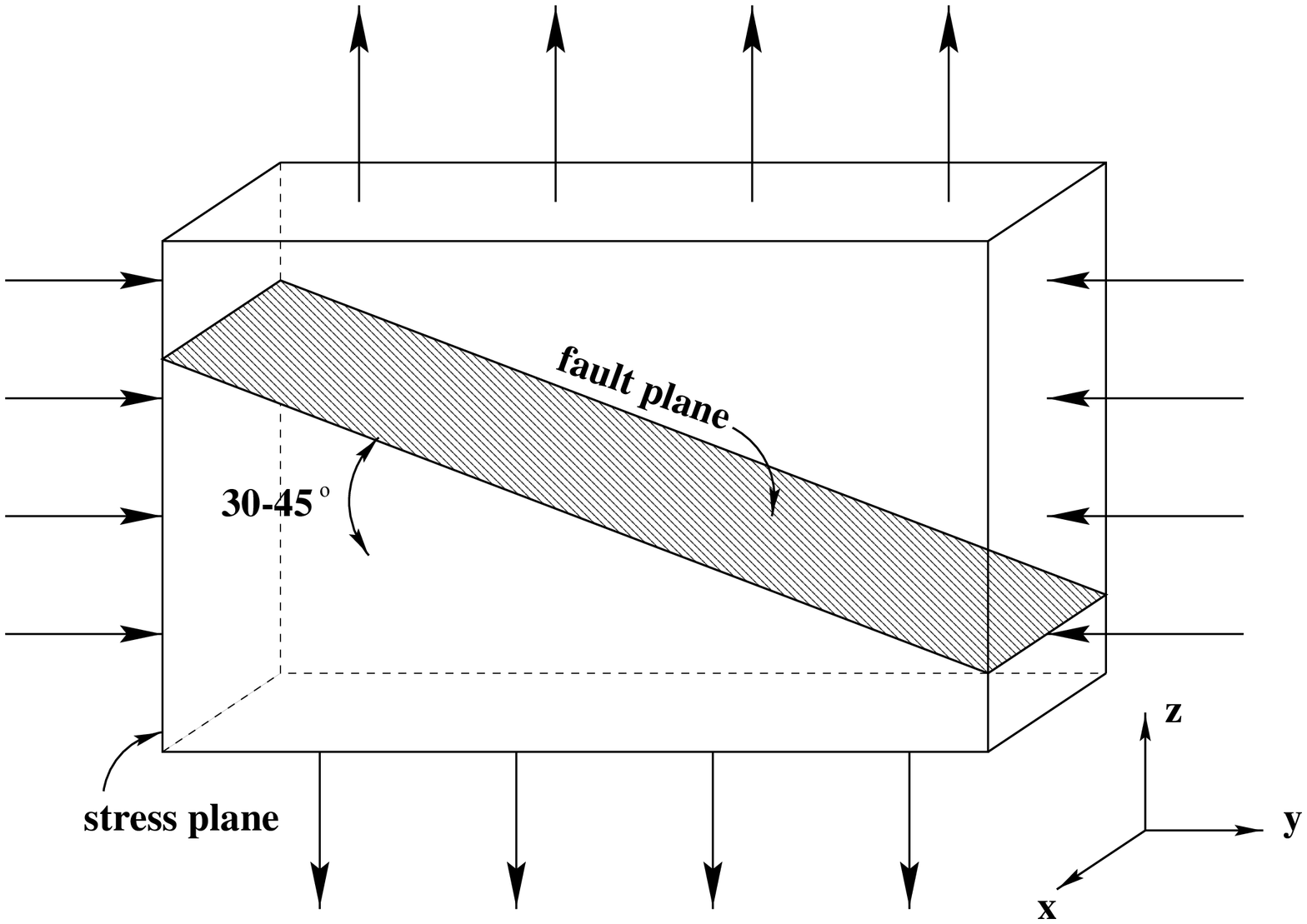}
\figcaption{Breaking of a compressed matter element. A block of matter
subjected to horizontal compression and vertical tension shears along a
fault plane as shown when critical strain is reached. Shearing along a
complementary plane flipped over with respect to the plane shown is
equally likely for isotropic material. The $y$--$z$ plane is the {\it
stress plane}.}
\end{figure}

\begin{figure}
\plotone{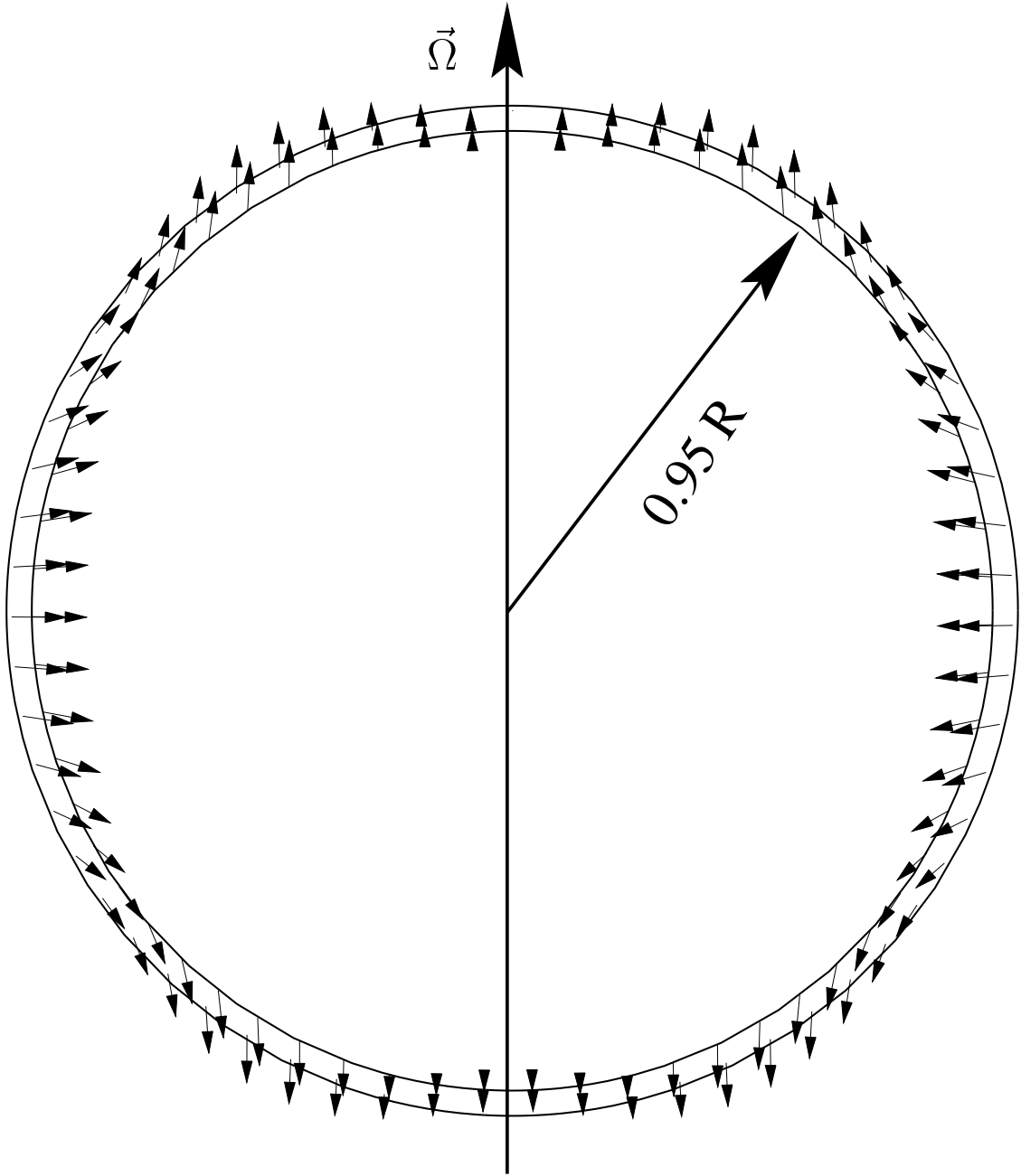}
\caption{The displacement field in a spinning-down neutron star. A
cross section through the center of the star is shown for a model with
relative crust thickness of 5\%. The equatorial diameter decreases while
the polar diameter increases.}
\end{figure}

\begin{figure}
\plotfiddle{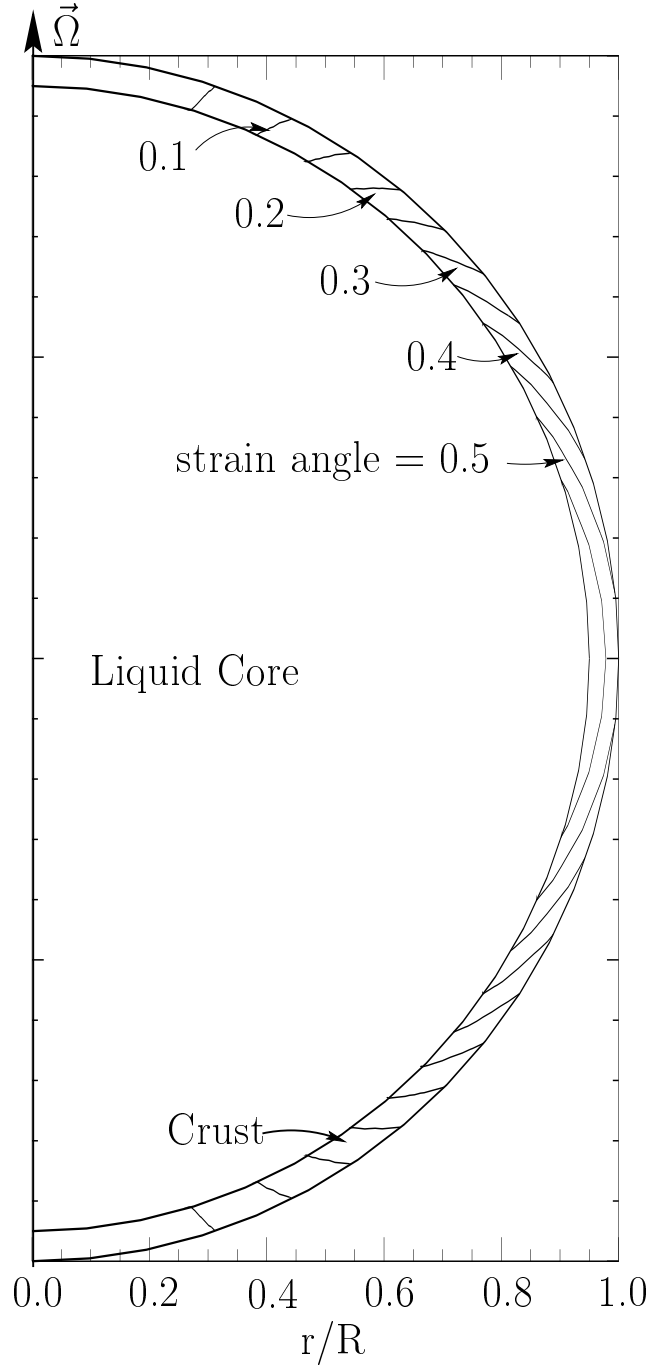}{7.0in}{0}{80}{80}{-200pt}{-50pt}
\figcaption{Contours of constant strain angle as a function of depth in
the crust for a non-magnetized neutron star (arbitrary units). The
maximum strain angle occurs at the equator and on the inner boundary of
the stellar crust.}
\end{figure}

\begin{figure}
\plotone{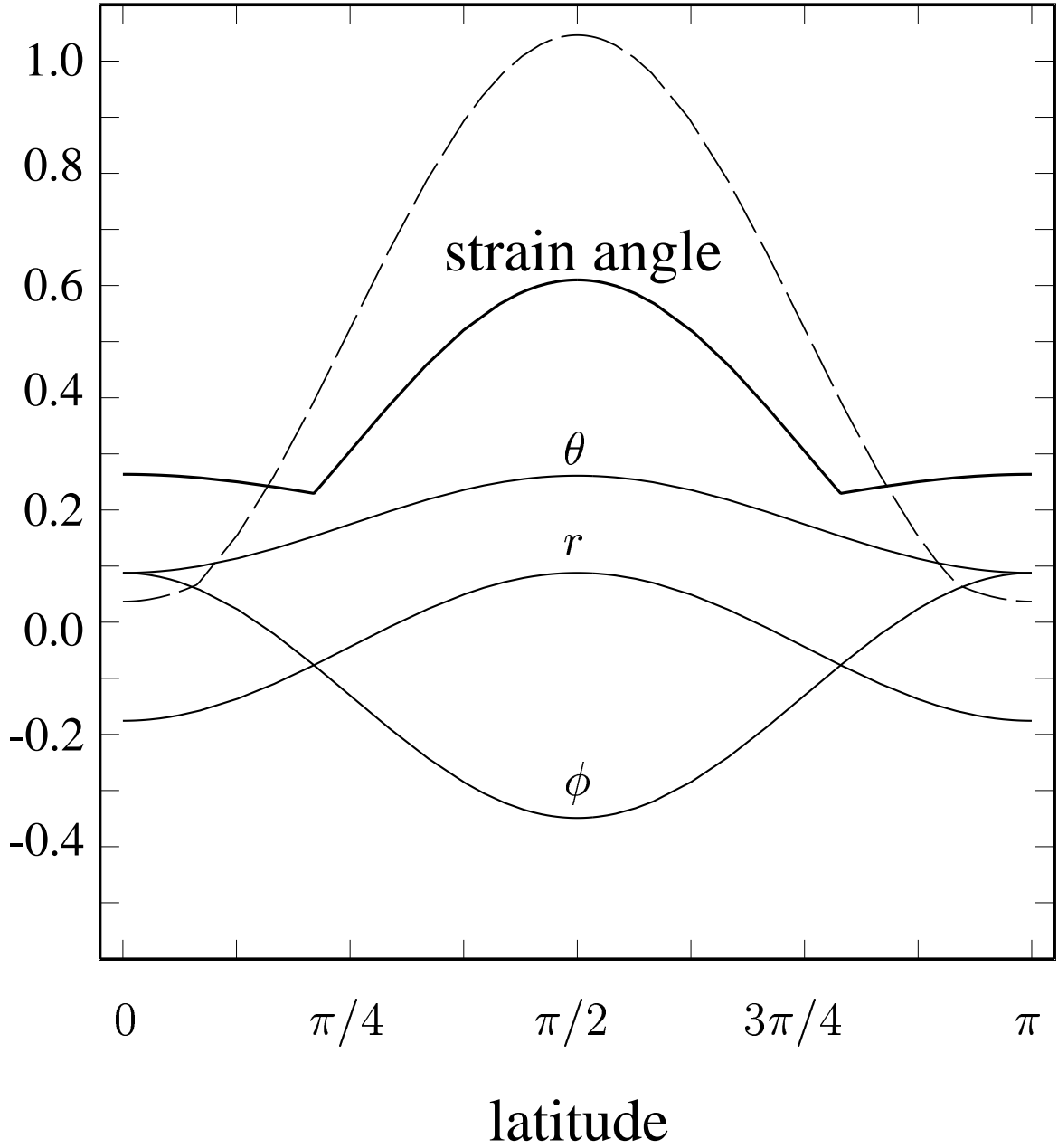}
\figcaption{Strain eigenvalues and strain angle vs. latitude at
the stellar surface for $B=0$ (arbitrary units). The curves
show the strain angle and the eigenvalues as marked. The dashed line
corresponds to the strain angle for the homogeneous sphere of LFE.}
\end{figure}

\begin{figure}
\plotone{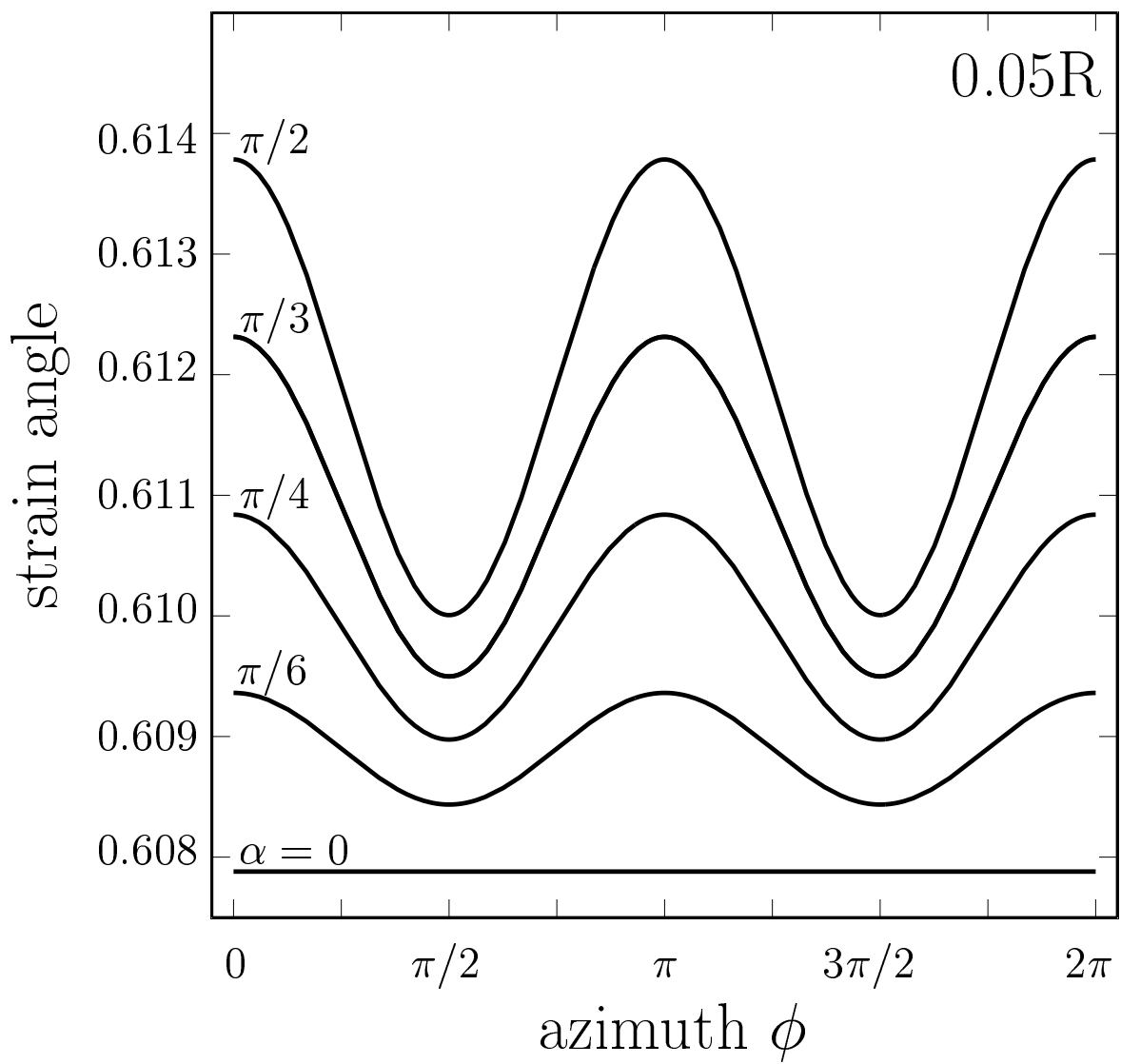}
\figcaption{The strain angle of surface equatorial material in the
presence of a magnetic field (arbitrary units). The curves correspond to
different values of the angle $\alpha$ between the magnetic and rotation
axes. The magnetic poles are at azimuthal angles $\pi/2$ and $3\pi/2$.}
\end{figure}

\begin{figure}
\plotfiddle{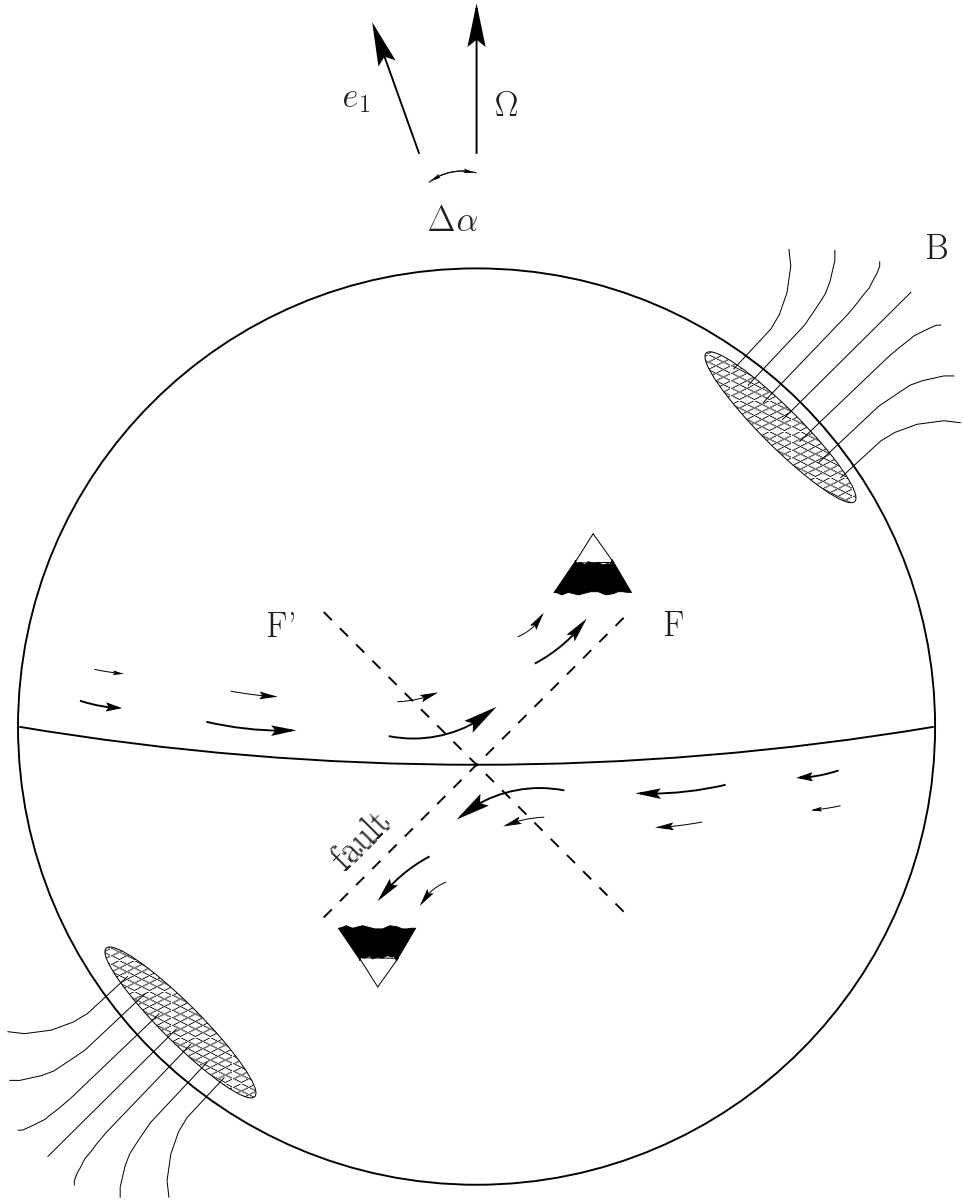}{6.0in}{0}{80}{80}{-250pt}{-100pt}
\figcaption{A starquake. In the absence of a magnetic field, the
material is equally likely to begin breaking along faults $F$ and
$F^\prime$, anywhere on the equator. These fault planes are
perpendicular to the stellar surface. In the presence of magnetic
stresses, fault $F$ is more likely, creating ``mountains'' (indicated
by the snow-capped peaks) and shifting the largest principal axis of
inertia to a new direction $e_1$ (fixed in the star).}
\end{figure}


\newpage

\begin{thebibliography}{}

\def\nature{{\rm Nature}}
\def\nucphys{{\rm Nuc. Phys.}}
\def\nucphysa{{\rm Nuc. Phys. A}}
\def\physletb{{\rm Phys. Lett. B}}
\def\physrevc{{\rm Phys. Rev. C}}
\def\physrevl{{\rm Phys. Rev. Lett.}}
\def\prd{{\rm Phys. Rev. D}}
\def\sovphysjetp{{\rm Soviet~Phys.~JETP}}
\def\ptpl{{\rm Prog. Theor. Phys. Lett.}}
\def\ptps{{\rm Prog. Theor. Phys. Suppl.}}
\def\ptp{{\rm Prog. Theor. Phys.}}

\bibitem[Akmal \etal\ 1998]{ns-eos}
Akmal, A., Pandharipande, V. R. \& Ravenhall, D. G. 1998, \physrevc, 58,
1804

\bibitem[Baym \&\ Pines 1971]{starquakes}
Baym, G. \&\ Pines, D. 1971, Ann. Phys., 66, 816

\bibitem[Blaes \etal\ 1989]{quake-grbs}
Blaes, O., Blandford, R., Goldreich, P. \& Madau, P. 1989, \apj, 343, 839

\bibitem{chandra} 
Chandrasekhar, S. 1987, Ellipsoidal Figures of Equilibrium
(New York, NY: Dover)

\bibitem[Cheng \etal\ 1996]{cheng} 
Cheng, B. L, Epstein, R. I., Guyer, R. A. \& Young, A. C. 1996,
\nature, 382, 518-520

\bibitem{dt} Duncan, R. C. \& Thompson, C. 1994, in {\sl Gamma-ray
Bursts}, Am. Inst. Phys., New York, (eds: Fishman, G. J., Brainerd,
J. J., Hurley, K.), p. 625

\bibitem[Gogus \etal\ 1999]{1900+14}Gogus, E., Woods, P. E.,
Kouveliotou, C., van Paradjis, J., Briggs, M. S., Duncan, R. C., \&
Thompson, C. 1999, \apj, in press (astro-ph/9910062)

\bibitem[Green \&\ Houston 1995]{gh95}Green, H. W. II \&\ Houston,
H. 1995, {\sl Ann. Rev. Earth Planet Sci.}, 23, 169

\bibitem[Hertzberg 1996]{hertzberg96}Hertzberg, R. W. 1996, Deformation
and Fracture Mechanics of Engineering Materials, (New York: Wiley)

\bibitem[Landau \&\ Lifshitz 1959]{ll}
 Landau, L. D. \& Lifshitz, E. M. 1959, Theory of Elasticity,
(London: Pergamon Press)

\bibitem[Link \& Epstein 1996]{thermal-glitches}
 Link, B., and Epstein, R. I. 1996, \apj, 457, 844

\bibitem[Link \& Epstein 1997]{le97}
 Link, B., and Epstein, R. I. 1997, \apj, 478, L91

\bibitem[LFE]{lfe98}Link, B., Franco, L. M.
\&  Epstein, R. I. 1998, \apj, 508, 838

\bibitem[Lorenz \etal\ 1993]{ns-eos-crust}
Lorenz, C. P., Ravenhall, D. G. \& Pethick, C. J. 1993, \physrevl, 70, 379

\bibitem[Lyne, Pritchard, \&\ Smith 1993]{LPS} Lyne, A. G., Pritchard,
R. S., \&\ Smith, F. G. Smith 1993, \mnras, 265, 1003

\bibitem[Ruderman 1976]{ruderman76}Ruderman, M. 1976, \apj, 203, 213

\bibitem{}
Shapiro, S. L. \& Teukolsky, S. A. 1983, Black Holes, White
Dwarfs, and Neutron Stars, (New York: Wiley)

\bibitem[Shemar \&\ Lyne 1996]{sl96}Shemar, S. L. \& Lyne, A. G. 1996,
\mnras, 282, 677

\bibitem[Strohmayer \etal\ 1991]{shear-mod}
Strohmayer, T., Ogata, S., Iyetomi, H., Ichimaru, S., \& Van Horn,
H. M. 1991, \apj, 375, 679

\bibitem{tang} Tang, A. 1999, M.S. Thesis, University of Hong Kong

\bibitem[Tauris 1998]{tauris}
Tauris, T. M., \& Manchester, R. N. 1998, \mnras, 298, 625

\bibitem[Thompson \& Blaes 1998]{tb}Thompson, C. \& Blaes, O. 1998,
\prd, 57, 3219

\bibitem{td95} Thompson, C. \& Duncan, R. C. 1995, 
\mnras, 275, 255
 
\bibitem[Thompson \& Duncan 1996]{td}Thompson, C. \& Duncan, R. C.
1996, \apj, 473, 322

\end{thebibliography}
\end{document}